\documentclass[aps,prd]{revtex4}
\usepackage{amsmath}
\usepackage{amssymb}

 \newread\testifexists
\def\GetIfExists #1 {\immediate\openin\testifexists=#1
    \ifeof\testifexists\immediate\closein\testifexists\else
    \immediate\closein\testifexists\input #1\fi}

\immediate\openin\testifexists=e
    \ifeof\testifexists\immediate\closein\testifexists\else
    \immediate\closein\testifexists\input e\fipsf

\begin{document}

\def\be{\begin{eqnarray}}
\def\en{\end{eqnarray}}
\def\non{\nonumber}
\def\la{\langle}
\def\ra{\rangle}
\def\pp{{\prime\prime}}
\def\nc{N_c^{\rm eff}}
\def\vp{\varepsilon}
\def\hep{\hat{\varepsilon}}
\def\a{{\cal A}}
\def\B{{\cal B}}
\def\c{{\cal C}}
\def\d{{\cal D}}
\def\e{{\cal E}}
\def\p{{\cal P}}
\def\t{{\cal T}}
\def\B{{\cal B}}
\def\P{{\cal P}}
\def\S{{\cal S}}
\def\T{{\cal T}}
\def\C{{\cal C}}
\def\A{{\cal A}}
\def\E{{\cal E}}
\def\V{{\cal V}}
\def\lr{\buildrel\leftrightarrow\over\partial}
\def\epu{\epsilon^{\mu\nu\alpha\beta}}
\def\epd{\epsilon_{\mu\nu\alpha\beta}}
\def\CP{$CP$~}
\def\up{\uparrow}
\def\dw{\downarrow}
\def\vma{{_{V-A}}}
\def\vpa{{_{V+A}}}
\def\smp{{_{S-P}}}
\def\spp{{_{S+P}}}
\def\lrpartial{\buildrel\leftrightarrow\over\partial}
\def\J{{J/\psi}}
\def\3bar{{\bf \bar 3}}
\def\6bar{{\bf \bar 6}}
\def\10bar{{\bf \ov{10}}}
\def\ov{\overline}
\def\Lqcd{{\Lambda_{\rm QCD}}}
\def\pr{{Phys. Rev.}~}
\def\prl{{ Phys. Rev. Lett.}~}
\def\pl{{ Phys. Lett.}~}
\def\np{{ Nucl. Phys.}~}
\def\zp{{ Z. Phys.}~}
\def\lsim{ {\ \lower-1.2pt\vbox{\hbox{\rlap{$<$}\lower5pt\vbox{\hbox{$\sim$}
}}}\ } }
\def\gsim{ {\ \lower-1.2pt\vbox{\hbox{\rlap{$>$}\lower5pt\vbox{\hbox{$\sim$}
}}}\ } }

\title{ Final State Interaction In $B\to KK$ Decays}

\author{Cai-Dian L\"u }
\address{\it \small  CCAST (World Laboratory), P.O. Box 8730,
   Beijing 100080, P.R. China\\
\it \small   Institute of High Energy Physics, CAS, P.O.Box
918(4), \it \small 100049, P.R. China\footnote {Mailing address}}
\author{Yue-Long  Shen \footnote {shenyl@mail.ihep.ac.cn}\,and
Wei Wang \footnote {wwang@mail.ihep.ac.cn} } \address{\it \small
Institute of High Energy Physics, CAS, P.O.Box 918(4),  \it \small
100049, P.R. China\\ \it \small   Graduate School of Chinese
Academy of Science, P.R. China }

\begin{abstract}
We study the final state interaction effects in $B\to KK$ decays.
We find that the $t$ channel one-particle-exchange diagrams cannot
enhance the branching ratios of $\ov{B}^0\to K^0\ov{K^0}$ and
$B^-\to K^0K^-$ very sizably. For the pure annihilation process
$\ov{B}^0\to K^+K^-$, the obtained branching ratio by final state
interaction is at ${\cal{O}}(10^{-8})$.
\end{abstract}
\maketitle

\section {Introduction}
$B$ meson non-leptonic decays are important to study CP violation
and to extract CKM parameters. When the $B$ meson decays into two
light mesons, the final state particles are energetic, so it is
argued that they do not have enough time to get involved in soft
final state interaction(FSI). In spite of the FSI, several
factorization approaches, such as the naive factorization approach
(FA) \cite{fac,akl1,chengfac}, the QCD factorization approach
(QCDF) \cite{bene}, the perturbative QCD approach (PQCD)
\cite{kls,lucd} and Soft-Collinear-Effective-Theory (SCET)
\cite{scet} have been established to analyze $B$ meson decays.
These approaches successfully explain many phenomenons, but there
are still some problems hard to explain within these frameworks,
which have been summarized in \cite{cheng}. These may be hints of
the need of FSI in $B$ decays. It has been argued that the FSI is
power suppressed for the cancellation of the various intermediate
states in the heavy quark limit \cite{bene}, but for the finite
bottom quark mass, this effect may not be very effective
\cite{buras}. So FSI may be important to the channels that are
suppressed by other factors (such as the color factor or the CKM
matrix elements). For example, $B\to KK$ decays are usually
considered to be in the category \cite{lhn}.

FSI effects are nonperturbative in nature, so it is difficult to
study in a systematic way and some different mechanism of the
rescattering effects have been considered.  In the study of $D$
meson decays, the form factors are introduced to parameterize the
offshellness of the exchanged particles \cite{form1, form2}, and
this method still works in $B$ meson case. This mechanism has been
used to explain some puzzles \cite{cheng, cheng2}, such as $B \to
\pi\pi, \pi K$ puzzle, it is argued that these puzzles can be
resolved by FSI if we adopt appropriate parameters. If this is the
right method to resolve these puzzles,
 it should be consistent with other channels, such as the small branching ratio
 of $B\to KK$ and $B\to \rho^0\rho^0$ decays. The $B\to KK$
 decays have been measured by Belle \cite {belle} and Babar \cite{babar}
 , which are shown in TABLE 1( where the world average values are taken from \cite{hfag}).
 The FA predictions can be consistent with the experiment for
 $B^0\to K^0 \ov{K}^0$ and  $B^+\to \ov{K}^0K^+$ if we employ
 the current nonperturbative
inputs \cite{akl1,bene}, thus the FSI effects may not be too
large. The $\ov{B}^0\to K^+{K^-}$ is a pure annihilation decay
channel, so it is expected to be very small in FA, and the FSI can
give sizable corrections. In this paper we will follow the method
in \cite{cheng}, focusing on the two body intermediate states and
considering only $t$-channel one-particle-exchange processes at
hadron level. We will give the detailed calculation of the FSI
effects for $B\to KK$ decays in the next section, and then a brief
summary in the third section.
\[
\begin{tabular}{c|c|c|c}
\multicolumn{4}{c}{TABLE.{1}.\ \ Measured  branching fractions ($\times 10^{-6}$) of $B\to KK$ decays} \\
\hline \hline
Channel& Babar & Belle & World average \\
\hline \hline  $B^0 \to K^0 \ov{K^0}$ & $1.19^{+0.40}
_{-0.35}\pm 0.13$ & $0.8 \pm 0.3 \pm 0.1$ & $0.96^{+0.25}_{-0.24 }$ \\
$B^0 \to K^+ {K^-}$ &$<0.6$
&$<0.37$& \\
$B^+ \to \ov{K^0}
K^+$ & $1.5 \pm 0.5 \pm 0.1$&$1.0 \pm 0.4 \pm 0.1$ & $1.2 \pm 0.3$\\
\hline
\end{tabular}\]

\section{Final State Interactions Effects In $B\to KK$ Decays}

Before analyzing the FSI in $B\to KK$ decays, we first explore
what we can get in the usual short distance analysis. The short
distance contribution of the heavy meson decays can be expressed
in terms of some types of quark diagrams:  $\P$, the penguin
emission diagram; $\E$, W-exchange diagram; $\A$, $
W$-annihilation diagram; $\P_A$, the penguin annihilation diagram
(space-like); $\P_{EW}$, the electroweak penguin diagram; $\V$,
the vertical $W$ loop diagram (time-like penguin). The penguin
dominated $B\to KK$ decays can be expressed as: \be \non
 &&A(\ov{B}^0\to K^0\ov{K^0})=\P+\P_A-\frac{1}{3}\P_{EW}+\V, \\ \non
 &&A(B^-\to K^0K^-)=\P+\P_A-\frac{1}{3}\P_{EW}+\A, \\
 &&A(\ov{B}^0\to K^+K^-)=\E+\V.
\en
In factorization approach, there is no emission tree diagram
contribution to these decays. The annihilation diagrams
$\A,\E,\V,\P_A$ are power suppressed which can be neglected in the
calculation. They are usually believed to be long distance
dominant. So the short distance amplitudes read:
 \be
  A(\ov{B}^0\to K^0\ov{K^0})&&=
 i\frac{G_F}{\sqrt{2}}f_KF_0^{BK}(m_K^2)(m_B^2-m_K^2)[V_{ub}V_{ud}^*(a_4^u+r^K_{\chi}a_6^u)
 +V_{cb}V_{cd}^*(a_4^c+r^K_{\chi}a_6^c)\non\\  &&+V_{ub}V_{ud}^*(a_{10}^u+r^K_{\chi}a_8^u)
 +V_{cb}V_{cd}^*(a_{10}^c+r^K_{\chi}a_8^c)], \en  and $A(B^-\to
 K^0K^-)=A(\ov{B}^0\to K^0\ov{K^0})$
, $A(\ov{B}^0\to K^+K^-)=0$,
 where $ V_{ub},V_{ud},V_{cb}$ and $V_{cd}$ are CKM matrix elments, $r^K_{\chi}=2m_K^2/[m_b(m_s+m_q)]$.
 $a_i^{u,c}$ are combination of Wilson coefficients for four quark operators defined in in ref.\cite{akl1}.
 \be
 \non
 &&a_i=C_i+\frac{1}{3}C_{i+1},\,\,(i=odd)\\
 &&a_i=C_i+\frac{1}{3}C_{i-1},\,\,(i=even)
 \en
 From quark-hadron duality, the decay amplitude
 can be got from either quark picture or hadron picture. The result should be equal. However,
 neither of the two pictures are fully understood in the B
decays. The factorization theorem
 tells us to calculate the short distance contribution
 perturbatively and the long distance parts using hadronic picture. Thus a double counting problem may arise.
 To avoid
 double counting, we adopt leading order Wilson coefficient at the
 scale $m_b$ for naive factorization approach instead of QCDF (which includes some virtual corrections from long distance)
   for short distance calculations of $B\to KK$.

When we calculate the long distance contributions to the decays,
we consider only the CKM most favored two body intermediate
states, such as $D^{(*)}D^{(*)}, \pi\pi, \rho\rho$. The quark
level $B\to \pi\pi(\rho\rho)\to KK$ diagrams are shown in Figure
1. We can see that the this diagram has the same topology as the
penguin diagram or $W$-exchange diagram. From Eq.(1), we can see
that this kind of diagrams can contribute to $B\to
K^0\ov{K^0},K^+K^-, K^0K^-$ simultaneously. When the intermediate
state is $D^{(*)+}D^{(*)-}(D^{(*)+}\bar{D}^{(*)0})$, only penguin
topology works, so it cannot contribute to the $\ov{B}^0\to
K^+K^-$ decay.
\begin{figure}[tbh]
\begin{center}
\epsfxsize=5.0in\leavevmode\epsfbox{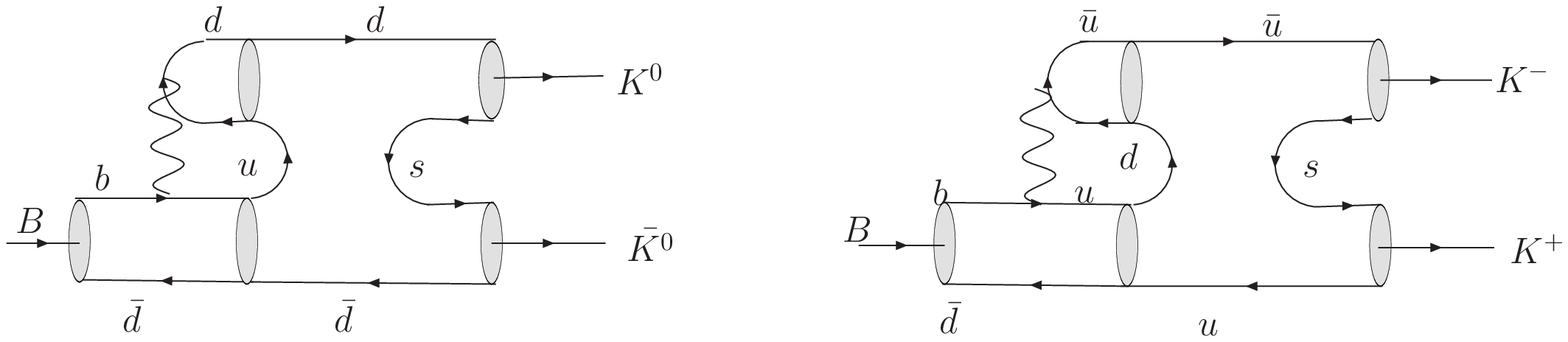}
\end{center}
\caption{{\protect\footnotesize Quark level diagram for $B\to
\pi^+\pi^-\to K^0\ov{K^0}(K^+K^-)$}}
\end{figure}

The hadron level diagrams are given in Figure 2. We focus on the
$t$ channel one-particle-exchange processes, furthermore, we
consider only the case that the two intermediate particles are on
shell, i.e. we only keep the absorptive part of diagrams in Figure
2, which gives the main contribution.
\begin{figure}[tbh]
\begin{center}
\epsfxsize=4.0in\leavevmode\epsfbox{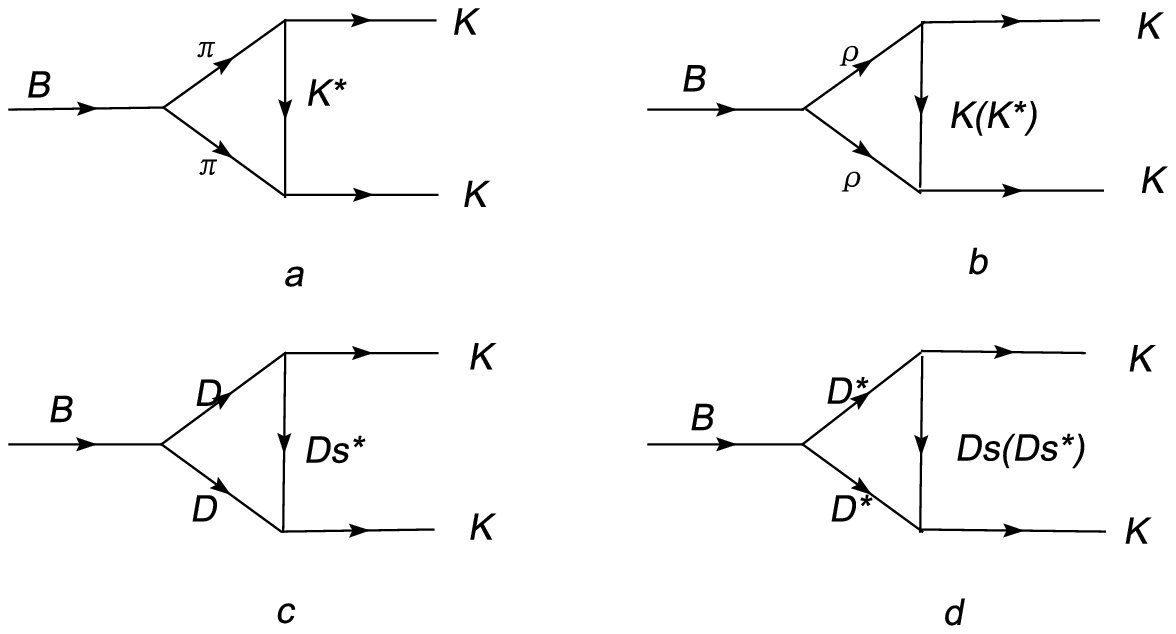}
\end{center}
\caption{{\protect\footnotesize Hadron level diagrams for long
distance $t$ channel contribution to $B\to KK$}}
\end{figure}

The absorptive part of the diagrams in Figure 2 can be calculated
with the following formula: \be \non \a bs\,A(P_B\to p_3p_4)= &&
{1\over 2}\int {d^3 p_1\over (2\pi)^32E_1}\,{d^3
 p_2\over
(2\pi)^3 2E_2}\,(2\pi)^4\delta^4(p_3+p_4-p_1-p_2)A(P_B\to
p_1p_2)\\ \times &&T^*(p_3p_4\to p_1p_2), \label{eq:abs(a)}  \en
 which can be deduced using the optical theorem \cite{cheng}.

 Taken FSI corrections into account, the topological amplitudes are:
 \be
&& \non \P=\P_{SD}+i\A bs(a+b+c+d),\\
&& \E=i \A bs(a+b).
 \en
 Then the decay amplitudes turn to:\be
&&\nonumber A(\ov{B}^0\to K^0\ov{K^0})=\P+\P_{EW}+i\A bs(a+b+c+d),
\\ \non
 &&A(B^-\to K^0K^-)=\P+\P_{EW}+i\A bs(a+b+c+d), \\
 &&A(\ov{B}^0\to K^+K^-)=i\A bs(a+b).
 \en
 To perform the calculation, we introduce the relevant Lagrangian
 density \cite{Casalbuoni}:
 \be
 { \cal L} _l&=& -{1\over
4}Tr[F_{\mu\nu}(V)F^{\mu\nu}(V)]+ig_{VPP}Tr(V^\mu
P\buildrel\leftrightarrow\over\partial_{\mu}
P)+g_{VVP}\epsilon^{\mu\nu\alpha\beta}Tr(\partial_\mu V_\nu
\partial_\alpha V_\beta P),\label{lagrangian1}\\
 \non { \cal L}_D&=&-ig_{D^*DP}(D^i\partial^{\mu}P_{ij}D^{*j\dagger}_{\mu}-
  D^{*i}_{\mu}\partial^{\mu}P_{ij}D^{j\dagger})
-\frac{1}{2}g_{D^*D^*P}\epd D_{i}^{*\mu}\partial^{\nu}P^{ij}
\buildrel\leftrightarrow\over{\partial^{\alpha}}D_j^{*\beta\dagger}\\
\non
&&-ig_{DDV}D_i^{\dagger}\buildrel\leftrightarrow\over\partial_{\mu}D^j
(V^{\mu})^i_j-2f_{D^*DV}\epd
(\partial^{\mu}V^{\nu})^i_j(D_i^{\dagger}\buildrel\leftrightarrow\over{\partial^{\alpha}}D^{*\beta
j}-D^{*\beta\dagger} _i
\buildrel\leftrightarrow\over{\partial^{\alpha}} D^j)\\
&&+ig_{D^*D^*V}D^{*\nu\dagger}_i\buildrel\leftrightarrow\over\partial_{\mu}D_{\nu}(V^{\mu})^i_j
+4if_{D^*D^*V}D^{*\dagger}_{i\mu}(\partial^{\mu}V^{\nu}-\partial^{\nu}V^{\mu})D^{*j}_{\nu},
 \label{lagrangian2}\en
 where $P$ and $V_{\mu}$ are pseudoscalar and vector multiplets
 respectively. Here we take the convention $\epsilon^{0123}=1$.

 Using Eq. (\ref{eq:abs(a)}) and the
 Feynman rules derived from the Eqs. (\ref{lagrangian1}) and (\ref{lagrangian2}), we can get
 the leading long distance rescattering amplitude:
 \be
 \a bs(a)
 =\int_{-1}^1\ {|{\bf p}_1|d\cos\theta\over 16\pi m_B}\,{g^2_{K^*K\pi}}\,A(\ov
B^0\to \pi ^+\pi ^-)\,{F^2(t,m_{K^*})\over
 {t-m_{K^*}^2}+im_{K^*}\Gamma_{K^*}}\,H_1,
 \en
 with
 \be
  A(\ov{B}^0\to \pi^+\pi^-)&=&
 i\frac{G_F}{\sqrt{2}}f_\pi F_0^{B\pi}(m_\pi^2)(m_B^2-m_\pi^2)[V_{ub}V_{ud}^*(a_1+a_4^u+a_{10}^u+r^\pi_{\chi}(a_6^u+a_8^u))\non\\
 &+&V_{cb}V_{cd}^*(a_4^c+a_{10}^c+r^\pi_{\chi}(a_6^c+a_8^c))],\non\\
H_1&=&-(p_1\cdot p_2+p_3\cdot p_4+p_1\cdot p_4+p_2\cdot
p_3)-{(m_1^2-m_3^2)(m_2^2-m_4^2)\over m_{K^*}^2},
 \en
 where we denote the momentum by $B(p_B)\to \pi(p_1)\pi(p_2)\to K(p_3)K(p_4)$,
  $\theta $ is the angle between ${\bf p}_1$ and ${\bf p}_3$, and $r_\chi^\pi=2 m_\pi^2/[m_b(m_u+m_d)]$.
  Here $F(t,m_{K^*})$ is the form factor introduced to denote
 offshellness  of the exchanged particle, which is usually
 parameterized as \cite{cheng}:
 \be
 F(t,m)=(\frac{\Lambda^2-m^2}{\Lambda^2-t})^n.
 \en
 It is normalized to unity at $t=m^2$ ( $t$
 is the invariant mass of the exchanged particle), where we usually take $n=1$. The cutoff
 $\Lambda$ should not be far from the physical mass of the
 exchanged particle, where we choose
 \be
 \Lambda=m_{exc}+\eta \Lambda_{QCD}.
 \en
 The parameter $\eta$ depends not only on exchanged particle,
 but also on the external particles involved in the strong
 interaction. If it is determined from the $B\to \pi\pi$ branching
 ratios, then we can employ it in $B \to KK$ decays for $SU(3)$
 symmetry.

 Likewise, the absorptive parts of the other diagrams are given by
\be
 \a bs(b(K))
&=& -i{{G_F}\over{\sqrt2}}\,V_{ub}V_{ud}^*\int_{-1}^1\ {|{\bf
p}_1|d\cos\theta\over 16\pi
  m_B}\,4g_{\rho KK}^2\,{F^2(t,m_K)\over t-m_K^2}\,\non \\
  &\times& f_\rho
  m_\rho\,\Big[(m_B+m_\rho)A_1^{B\rho}(m_\rho^2)H_2-{2A_2^{B\rho}(m_\rho^2)\over
 (m_B+m_\rho)}H'_2\Big],\non\\
 \a bs(b(K^*))
&=& i{G_F\over{\sqrt2}}\,V_{ub}V_{ud}^*\int_{-1}^1{|{\bf
p}_1|d\cos\theta\over 16\pi m_B}\,g_{\rho K^*
K}^2\,{F^2(t,m_{K^*})\over
t-m^2_{K^*}+im_{K^*}\Gamma_{K^*}}\,\non\\
 &\times & f_\rho m_\rho\,\Big[(m_B+m_\rho)A_1^{B\rho}(m_\rho^2)H_3-{2A_2^{B\rho}(m_\rho^2)\over
(m_B+m_\rho)}H'_3\Big],\non\\
  \a bs(c)
 &=&\int_{-1}^1\ {|{\bf p}_1|d\cos\theta\over 16\pi m_B}\,{g^2_{D^*_sDK}}\,A(\ov
B^0\to D^+ D^-)\,{F^2(t,m_{D^*_s})\over
 {t-m_{D^*_s}^2}}\,H_4,\non \\
 \non \a bs(d(D_s))
 &=&-i{G_F\over\sqrt 2}V_{cb}V_{cd}^*\int_{-1}^1 \frac{|{\bf p}_1|d\cos\theta}{ 16\pi
 m_B}{g^2_{D_sD^*K}}\frac{F^2(t,m_{D_s})}
 {t-m_{D_s}^2}\,\non\\
 &\times& f_{D^*} m_{D^*}
 \Big[(m_B+m_{D^*})A_1^{BD^*}(m_{D^*}^2)H_5-\frac{2A_2^{BD^*}(m_{D^*}^2)}
 {m_B+m_{D^*}}H'_5 \Big],\non \\
\a bs(d(D_s^*))&=&
  i\frac{G_F}{\sqrt{2}}V_{cb}V_{cd}^*\int_{-1}^1 {|{\bf p}_1|d\cos\theta\over 16\pi m_B}g^2_{D_s^*D^*K}
  \frac{F^2(t,m_{D^*_s})} {t-m_{D^*_s}^2}\non
  \\
   &\times& f_{D^*}m_{D^*}\Big[(m_B+m_{D^*})A_1^{BD^*}(m_{D^*}^2)H_6-\frac{2A_2^{BD^*}(m_{D^*}^2)}
 {m_B+m_{D^*}}H'_6\Big],
 \en

where
 \be
 H_2 &=&(p_3\cdot p_4)-{p_1\cdot p_3 p_1\cdot p_4 \over m_1^2}-{p_2\cdot p_3 p_2\cdot p_4 \over
 m_2^2}+{p_1\cdot p_2 p_1\cdot p_3 p_2\cdot p_4 \over
 m_1^2m_2^2},\non \\
 H'_2 &=& (p_3\cdot p_B)(p_4\cdot p_B)-{(p_1\cdot p_3)(p_1\cdot p_B)(p_4\cdot p_B)\over
 m_1^2}-{(p_2\cdot p_4)(p_2\cdot p_B)(p_3\cdot p_B)\over
 m_2^2}\non \\
 &&+{(p_1\cdot p_3)(p_2\cdot p_4)(p_1\cdot p_B)(p_2\cdot p_B)\over
 {m_1^2 m_2^2}},\non\\
  H_3&=&2(p_1\cdot p_4)(p_2\cdot
p_3)-2(p_1\cdot p_2)(p_3\cdot p_4),\non \\
 H'_3&=&m_B^2[(p_1\cdot p_4)(p_2\cdot p_3)-(p_1\cdot p_2)(p_3\cdot p_4)]
 +(p_1\cdot
 p_B)(p_2\cdot p_B)(p_3\cdot p_4)\non \\
 && -(p_2\cdot p_B)(p_3\cdot
 p_B)(p_1\cdot p_4)-(p_1\cdot p_B)(p_4\cdot p_B)(p_2\cdot
 p_3)+(p_3\cdot p_B)(p_4\cdot p_B)(p_1\cdot p_2),\non\\
  H_4&=&-(p_3\cdot p_4)+{(p_1\cdot p_3-m^2_3)(m^2_4-p_2\cdot
p_4)\over m_{D^*}^2},\non\\
 H_5&=&(p_3\cdot p_4)-{(p_1\cdot p_3)(p_1\cdot p_4)\over m_1^2}-{(p_2\cdot p_3)(p_2\cdot p_4)\over
 m_2^2}+{(p_1\cdot p_3)(p_2\cdot p_4)(p_1\cdot p_2)\over {m_1^2
 m_2^2}}, \non \\
 H'_5&=&(p_3\cdot p_B)(p_4\cdot p_B)-{(p_1\cdot p_B)(p_4\cdot
 p_B)(p_1\cdot p_3)\over m^2_1} \non \\
 && -{(p_2\cdot p_B)(p_3\cdot
 p_B)(p_2\cdot p_4)\over m^2_1}+{(p_1\cdot p_B)(p_2\cdot p_B)(p_1\cdot
 p_3)(p_2\cdot p_4)\over {m^2_1m^2_2}},\non\\
 H_6 &=& 2(p_1\cdot p_2)(p_3\cdot p_4)-2(p_1\cdot p_4)(p_2\cdot
 p_3), \non\\
 H'_6 &=& m^2_B[(p_1\cdot p_2)(p_3\cdot p_4)-(p_1\cdot
 p_4)(p_2\cdot p_3)]-(p_1\cdot p_B)(p_2\cdot p_B)(p_3\cdot p_4) \non\\
 & & +(p_2\cdot p_B)(p_3\cdot p_B)(p_1\cdot p_4)+(p_1\cdot
 p_B)(p_4\cdot p_B)(p_2\cdot p_3)-(p_3\cdot
 p_B)(p_4\cdot p_B)(p_1\cdot p_2).
 \en
 and
 \be
 A(\ov
B^0\to D^+ D^-)&=&
 i\frac{G_F}{\sqrt{2}}f_D F_0^{BD}(m_D^2)(m_B^2-m_D^2)[V_{cb}V_{cd}^*(a_1+a_4^c+a_{10}^c+m_D/m_B(a_6^u+a_8^u))\non\\
 &+&V_{ub}V_{ud}^*(a_4^u+a_{10}^u+m_D/m_B(a_6^u+a_8^u))]
 \en

 To proceed the numerical calculation, we use the parameters as
 follows: the Fermi constant
 $G_F = 1.166\times 10^{-5}GeV^{-2}$; the CKM matrix elements
$V_{cb} = 0.041, V_{cd} = -0.224, |V_{ub}| = 0.0037, V_{ud} =
0.974$; The phase angle $\gamma=60^{\circ}$; the meson and quark
masses $m_B = 5.279GeV, m_K = 0.498GeV, m_b = 4.4GeV, m_s =
0.09GeV, m_d = 0.004GeV$;
 the decay constants $f_\pi=0.132 GeV$, $f_D=0.20 GeV$,
 $f_\rho=0.216 GeV$, $f_{D^*}=0.23 GeV$, $f_K=0.16 GeV$;
 The form factors are from the light-front model \cite{CCH}:
 $F^{BK}(0)=0.35$, $A_1^{B\rho}(0)=0.22$, $A_2^{B\rho}(0)=0.20$, $F^{BD}(m_D^2)=0.68$,
  $A_1^{BD^*}(m_{D^*}^2)=0.65$. The
 coupling relevant to the $K^*K\pi$ can be extracted from the
 $K^* \to K\pi$ experiments: $g_{K^{*+}K^0\pi^+}$=4.6, and we take $g_{\rho KK}=4.28$
 and $g_{\rho K K^*}=8\sqrt{2}$ \cite{cheng}.
  The coupling of $D_s^*DK$ and
 $D_s^*D^*K$ can be related to $g_{D^*D\pi}$ by $SU(3)$ symmetry. In this work we neglect
  the SU(3)symmetry breaking effect
 and employ the coupling as $g_{D_s^*DK}=\sqrt{m_Dm_{D^*}}g_{D_s^*D^*K}=g_{D^*D\pi}=17.9$.
 Similarly, we also use the symmetry to determine the parameter
 $\eta$ in the form factor, where the best fit from the $B\to \pi K$ decay is
  $\eta_\rho = \eta_{D^{(*)}({D_s^{(*)}})}=0.69$ \cite{cheng}, in this work
 we choose $\eta=(0.8,1.0,1.2)\times 0.69$ to include the $SU(3)$ breaking effect.

 The rescattering effects can produce the strong phases, it may
 change the CP asymmetry behavior of short distance calculation.
 The time dependent CP asymmetry of $B^0 \to K^0\ov{K^0}$ is
 defined as
 \be
 A_{CP}({B^0(t)\to K^0\ov{K}^0})&& \nonumber =\frac{\Gamma(\ov{B}^0(t)\to K^0\ov{K}^0)-
 \Gamma(B^0(t)\to K^0\ov{K}^0)}{\Gamma(\ov{B}^0(t)\to K^0\ov{K}^0)+\Gamma(B^0(t)\to
 K^0\ov{K}^0)}\\
 && = A_{K^0\ov{K}^0}\cos(\Delta M t)+ S_{K^0\ov{K}^0}\sin(\Delta M
 t),
 \en
with $\Delta M$ the mass difference of the two mass eigenstates of
neutral mesons. And the direct CP asymmetry and the mixing induced
CP asymmetry parameters are defined as, \be A_{K^0\ov{K}^0}
=\frac{|\lambda_{K^0\ov{K}^0}|^2-1}{|\lambda_{K^0\ov{K}^0}|^2+1},\,\,\,\,\,\,\,
 S_{K^0\ov{K}^0}
=\frac{2\rm{Im}(\lambda_{K^0\ov{K}^0})}{|\lambda_{K^0\ov{K}^0}|^2+1},
\en where the corresponding factor
$\lambda_{K^0\ov{K}^0}=e^{-2i\beta}\frac{\bar{A}}{A}$.

   Using the theoretical inputs mentioned above, we get flavor-averaged branching ratios for the
   short distance contribution as
   \be
   \nonumber &&  {\cal B}(B^0\to K^0\ov{K}^0)=0.94\times
   10^{-6},\\
   &&  {\cal B}(B^+\to \ov{K}^0K^+)=1.0\times
   10^{-6}.
   \en
   And there is no direct $
   CP$ violation since there is only one kind of contribution (pure penguin). After considering rescattering
   effects, things will change, since more contributions with different phases are introduced. We summarize our
   numerical results in TABLE 2.
  \[
\begin{tabular}{c|c|c|c|c}
\multicolumn{5}{c}{TABLE {2}.\ \ CP averaged branching ratios and CP asymmetries of $B\to KK$ decays} \\
\hline \hline
Channel& $\eta(\times 0.69)$ & Branching ratio($\times 10^{-6}$) & $A_{KK}$ & $S_{KK}$\\
\hline \hline  &0.8&0.99 & -0.03 & -0.03 \\ $B^0 \to K^0 \ov{K}^0$
& 1.0&1.1 & -0.04 &-0.04
\\& 1.2&1.2 &-0.06 & -0.05   \\
\hline & 0.8&0.009 & -0.04 & -0.56  \\ $B^0 \to K^+
{K^-}$ & 1.0&0.021 & -0.04 &-0.55
\\& 1.2&0.042 &-0.03 & -0.55   \\
\hline & 0.8&1.1 & 0.10 & - \\ $B^+ \to \ov{K}^0 K^+$ & 1.0&1.2 &
0.14 &-
\\& 1.2&1.3 &0.18& - \\
\hline \end{tabular}\]

From this table, we can see that the FSI cannot enhance the
branching ratio of $B^0(\ov{B}^0)\to K^0\ov{K}^0$ sizably because
the FSI increase(decrease) the real part for $B^0 \to
K^0\ov{K}^0$($\ov{B}^0\to K^0\ov{K}^0$), but decrease(increase)
the imaginary part. The total effects don't make the average
branching ratio change much. As the parameter $\eta$ gets larger,
the FSI effects become more important and the larger strong phase
is produced, so the absolute value of direct and the mixing
induced asymmetry increases. For the charged $B$ meson decays, the
FSI effects are more important for Figure 2(a, b) give double
contribution (due to the interchange of the intermediate
particles). So contrary to $B^0 \to K^0\ov{K}^0$ case, the direct
$CP$ asymmetry becomes positive. The $B^0(\ov{ B^0}) \to K^+
{K^-}$ results are purely from the FSI effects, its branching
ratio are of the order ${\cal{O}}(10^{-8})$, which is consistent
with PQCD prediction \cite{lhn} in quark diagram calculation. It
seems to be a proof for quark hadron duality. The $D(D^*)D(D^*)$
intermediate states cannot contribute to $B^0(\ov {B^0}) \to K^+
{K^-}$ through $t$ channel processes, the strong phase of this
channel comes from the Wilson coefficients, so the calculation
gives a small direct CP asymmetry.

  In ref \cite{cheng}, the $D\ov {D} \to \pi\pi$ annihilation diagrams which
  have the same topology with vertical $W$ loop diagrams, are
  introduced to resolve $B\to\pi\pi$ puzzle. It gives an dispersive part which
   can reduce $B^0 \to
  \pi^+\pi^-$ branching ratio as well as enhance $B^0 \to
  \pi^0\pi^0 $ one. Considering $SU(3)$ symmetry, these
  diagrams can contribute to $B \to KK$ at the same level as $B \to
  \pi\pi$, we quote their results here (in units of $GeV$):
  $\rm{Dis} A= 1.5\times 10^{-6}V_{cb}V^*_{cd}-6.7\times 10^{-7}V_{ub}V^*_{ud}$.
  If we consider this effect in $B \to KK$ case, the
  branching ratio for $B\to K^+K^-$  is enhanced to about
  $2\times 10^{-6}$, while the $B^0 \to K^0\ov{K}^0$ branching ratio is
  reduced to about $6\times 10^{-7}$, which is not favored by $B \to
  KK$ experimental  data.

  The $B \to KK$ decays have also been calculated with the QCD
  factorization\cite{qcdf} and PQCD approach \cite{lhn}, in which
  part of the long-distance effects has been included. These
  methods depend strongly on theoretical inputs, such as the
  chiral factor(or equivalently, the current quark mass), so they
  also
  give large error. The QCDF calculations give (branching ratios are CP averaged, also for (\ref{pqcd})):
  \be
  \nonumber && {\cal B}(B^0 \to K^0\ov{K}^0)=1.35^{+0.41+0.70+0.13+1.09}_{-0.
  36-0.48-0.15-0.45}\times10^{-6},\\
  && \nonumber  {\cal B}(B^- \to K^0K^-)=1.36^{+0.45+0.72+0.14+0.91}_{-0.
  39-0.49-0.15-0.40}\times10^{-6},\\
  && \nonumber  {\cal B}(B^0 \to K^+K^-)=0.013^{+0.005+0.008+0.000+0.087}
_{-0.005-0.005-0.000-0.011}\times10^{-6},\\
  &&A_{CP}(B^- \to
  K^0K^-)=-16.3^{+4.7+5.0+1.6+11.3}_{-3.7-5.7-1.7-13.3}\times10^{-2},\\
 \nonumber  && A_{CP}(B^0 \to
  K^0\ov{K}^0)=-16.7^{+4.7+4.5+1.5+4.6}_{-3.7-5.1-1.7-3.6}\times10^{-2}.
  \en
  And the PQCD calculations give:
\be
  \nonumber && {\cal B}(B^0 \to K^0\ov{K}^0)=1.75\times10^{-6},\\
  && \nonumber  {\cal B}(B^- \to K^0K^-)=1.66\times10^{-6},\\
  && \nonumber  {\cal B}(B^0 \to K^+K^-)=0.046\times10^{-6},\\
  &&A_{CP}(B^- \to
  K^0K^-)=0.11,\label{pqcd}\\
 \nonumber  && A_{CP}(B^0 \to
  K^0\ov{K}^0)=0,\\
  \nonumber  && A_{CP}(B^0 \to
  K^+K^-)=0.29.
  \en
 For the branching ratio, with the error, all the
 calculations can be consistent. As for the CP asymmetry, PQCD
 and QCDF have opposite sign, our calculation is consistent
 with PQCD for $B^- \to K^0K^-$, while our results have the
 same sign with QCDF for $B^0 \to K^0\ov{K}^0$. More experimental
 data are needed to test these predictions.

 \section{summary}
In this paper we study the FSI effects in $B \to KK$ decays. We
find that if we consider only the dominant $t$ channel
one-particle-exchange diagrams, the FSI effects cannot change the
branching ratio of $B^0 \to K^0\ov{K}^0$ and $B^+( B^-) \to
\ov{K}^0 K^+(K^0K^-)$ sizably, which is consistent with the
current experimental data. We also predict the branching ratio of
the $B^0(\ov {B^0}) \to K^+ {K^-}$ at ${\cal{O}}(10^{-8})$ by
purely $t$ channel FSI, which is consistent with the PQCD
prediction. We also calculate the $CP$ asymmetry in the $B \to KK$
decays. We test the $D\bar{D}$ annihilation diagram (which is of
great importance to resolve $B\to\pi\pi$ puzzle in FSI)
contribution and find it not favored by $B \to KK$ data.
\section{acknowledgement}
We thank H. Y. Cheng, C. K. Chua, M.Z. Yang and Y. Li for helpful
discussions. C. D. L\"u thanks Hai-Yang Cheng and Hsiang-nan Li
for the warm hospitality during his visit at Academia Sinica,
Taipei.


\begin{thebibliography}{99}
\bibitem{fac}M. Wirbel, B. Stech, M. Bauer, Z. Phys. C29, 637 (1985);
 M. Bauer, B. Stech, M. Wirbel, Z. Phys. C34, 103 (1987);
L.-L. Chau, H.-Y. Cheng, W.K. Sze, H. Yao, B. Tseng, Phys. Rev.
D43, 2176 (1991), Erratum: D58, 019902 (1998).

\bibitem{akl1} A. Ali, G. Kramer and C.D. L\"u, Phys. Rev. D58, 094009
(1998); C.D. L\"u, Nucl. Phys. Proc. Suppl. 74, 227-230 (1999).

\bibitem{chengfac}Y.-H. Chen, H.-Y. Cheng, B. Tseng, K.-C. Yang,
 Phys. Rev. D60, 094014 (1999); H.~Y.~Cheng, K.~C.~Yang,
Phys.\ Rev.\ D {\bf 62}, 054029 (2000).
\bibitem{bene}M. Beneke, G. Buchalla, M. Neubert, C.T. Sachrajda,
Phys. Rev. Lett83, 1914 (1999); M. Beneke, G. Buchalla, M.
Neubert, C.T. Sachrajda, Nucl. Phys. B591, 313 (2000), M. Beneke,
G. Buchalla, M. Neubert, C.T. Sachrajda, Nucl. Phys.
B606,245(2001).
\bibitem{kls}
Y.~Y.~Keum, H.~n.~Li and A.~I.~Sanda, Phys.\ Lett.\ B {\bf 504}, 6
(2001); Phys. Rev. D63, 054008 (2001).

\bibitem{lucd}C.D. L\"u, K. Ukai, M.Z. Yang, Phys. Rev. D63, 074009
(2001); C.~D.~Lu and M.~Z.~Yang, Eur.\ Phys.\ J.\ C {\bf 23}, 275
(2002).
\bibitem{scet}C.W. Bauer, S. Fleming, and M. Luke, Phys. Rev. D
{63}, 014006 (2001), C.W. Bauer, S. Fleming, D. Pirjol, and I.W.
Stewart, Phys. Rev. D {63}, 114020 (2001); C.W. Bauer and I.W.
Stewart, Phys. Lett. B {516} 134 (2001), Phys. Rev. D {65} 054022
(2002).
\bibitem{cheng}Hai-Yang  Cheng, Chun-Khiang Chua and Amarjit Soni, Phys. Rev. D71,
014030 (2005).
\bibitem{buras}A. T. Buras et al, Phys. Rev. Lett.92,
101804(2004), Nucl. Phys. B 697, 133(2004).
\bibitem{lhn} Chuan-Hung Chen and Hsiang-nan Li, Phys. Rev. D{\bf
63}, 014003 (2001).
\bibitem{form1} Y. Lu, B.S. Zou, and M.P. Locher,
\zp A {\bf 345}, 207 (1993); M.P. Locher, Y. Lu, and B.S. Zou,
{\it ibid} {\bf 347}, 281 (1994); X.Q. Li and B.S. Zou, \pl B {\bf
399}, 297 (1997); Y.S. Dai, D.S. Du, X.Q. Li, Z.T. Wei, and B.S.
Zou, \pr D {\bf 60}, 014014 (1999).

\bibitem{form2} M. Ablikim, D.S. Du, and M.Z. Yang, \pl B {\bf 536}, 34
(2002); J.W. Li, M.Z. Yang, and D.S. Du, hep-ph/0206154.
\bibitem{cheng2}Hai-Yang  Cheng, Chun-Kung Chua and Amarjit Soni, Phys.Rev. D72 (2005)
014006, arxiv: hep-ph/0502235.
\bibitem{belle} K. Abe et al., Phys. Rev. Lett95 (2005), 231802, arxiv: hep-ex/0506080.
\bibitem{babar} B. Aubert et al., Phys. Rev. Lett95 (2005), 221801,
arxiv: hep-ex/0507023.
\bibitem{hfag}Heavy Flavor Averaging Group, hep-ex/0505100; and
 references cited there.
\bibitem{Casalbuoni} R. Casalbuoni, A. Deandrea, N. Di Bartolomeo, R.
Gatto, F. Feruglio, and G. Nardulli, Phys. Rep. {\bf 281}, 145
(1997).
 \bibitem{CCH}
H.~Y.~Cheng, C.~K.~Chua, and C.~W.~Hwang, Phys.\ Rev.\ D {\bf 69},
074025 (2004).
\bibitem{qcdf} see, for example, M.
Beneke and M. Neubert, Nucl.Phys. B675 (2003) 333-415; D. S. Du,
H. J. Gong, J. F. Sun, D. S. Yang and G. H. Zhu, Phys. Rev. D64,
014036, (2001).

\end{thebibliography}
\end{document}